\documentclass{article}
\pdfoutput=1
\usepackage{spconf,amsmath,graphicx}
\usepackage{svg,comment}
\usepackage{adjustbox}
\usepackage{booktabs}
\usepackage{amsfonts}
\usepackage{color,url}
\usepackage[font=footnotesize]{caption}
\usepackage{multirow,multicol}

\title{Describing Emotions with acoustic property prompts for Speech Emotion Recognition}
\name{Hira Dhamyal$^1$, Benjamin Elizalde$^2$, Soham Deshmukh$^2$, Huaming Wang$^2$ , Bhiksha Raj$^1$, Rita Singh$^1$ \vspace{-8.5pt}
\thanks{This work was done when the first author was an intern at Microsoft}}
\address{$^1$ Carnegie Mellon University, $^2$Microsoft \\
\{hyd, bhiksha, rsingh\}@cs.cmu.edu, \{benjaminm, sdeshmukh, huawang\}@microsoft.com \vspace{-8pt}}
\begin{document}
%
\maketitle
\begin{abstract}

Emotions lie on a broad continuum and treating emotions as a discrete number of classes limits the ability of a model to capture the nuances in the continuum. The challenge is how to describe the nuances of emotions and how to enable a model to learn the descriptions. In this work, we devise a method to automatically create a description (or prompt) for a given audio by computing acoustic properties, such as pitch, loudness, speech rate, and articulation rate. We pair a prompt with its corresponding audio using 5 different emotion datasets. We trained a neural network model using these audio-text pairs. Then, we evaluate the model using one more dataset. We investigate how the model can learn to associate the audio with the descriptions, resulting in performance improvement of Speech Emotion Recognition and Speech Audio Retrieval. We expect our findings to motivate research describing the broad continuum of emotion.

\end{abstract}

\begin{keywords}
emotion recognition, contrastive language-audio pretraining, acoustic properties, prompt generation, prompt augmentation
\end{keywords}
\section{Introduction}
\label{sec:intro}
\vspace{-0.3cm}
Speech emotion recognition is the task of detecting emotion from a given audio. Emotion detection is playing an increasingly important role in the digital world today however there is still need for improvement. 
Humans express emotions on a very broad continuum. Models like the plutchik wheel of emotion \cite{plutchik1991emotions} or the Ekman's model of emotion \cite{ekman1992there} capture all emotions as a combination of 6 or 8 basic ones. Although these frameworks are extremely popular and provide ease of modelling, they limit the ability of machine learning models to capture the nuances in the spectrum of human emotions. 

The continuum of emotions instead of being categorized by handful of predefined classes, can be thought of in terms of some high dimensional continuous space, where any emotion can lie. 
This is important to model emotions since each expression of emotion is diverse and unique. It is dependent on speaker, culture, context among other factors. Labelling two instances of emotion with the same label of `anger', ignores the intricacy of the expression. Therefore, we explore modelling the continuity of emotions. 

This continuity of emotions can be captured by the flexibility that natural language descriptions provide. 
These descriptions can use affective language, that are often casually used to describe an emotion. Such affective language has acoustic correlates, for example: an angry man `shouting loudly' is describing the emotion by directly referring to the loudness or intensity of the speech. 

The choice of natural language description effects the high dimensional representation learned from the text, hence it is very important to choose the right description for the emotion. This leads to the question:

\noindent \emph{How do we describe an emotion using natural language and how can a model learn it?} 

In this work, we propose a method to describe the continuum of emotion by using the audio themselves to guide the descriptions. Previous research shows that there are numerous acoustic correlates of emotion \cite{frick1985communicating, scherer1972acoustic, pavlenko2005emotions}. These acoustic correlates  refer to the low level information like the 
average pitch, intensity, speech rate and articulation rate. We extract these correlates from each audio and use them to form the description in an automatic and scalable way. We call descriptions generated in this manner \emph{`acoustic prompts'}. 


Thus we need a model that learns to associate the audio and their descriptions. 
To do this, we build on top of the Contrastive Language-Audio Pretraining (CLAP) model \cite{elizalde2022clap}.
The model has separate audio and text encoders. It brings the audio and text representations to the same multimodal space. This architecture yields state of the art performance in learning audio concepts with natural language description. CLAP enables to evaluate the learned model on zero-shot and supervised classification and audio retrieval \cite{deshmukh2022audio}.

From our experiments, we find that acoustic prompts improve the model's performance in emotion classification settings.
Specifically, when the training dataset are relatively smaller, classification performance improves $3.8\%$ in Ravdess (Sec.~5.1). 
In a finetuning setup, we observe $3.7\%$ improvement (Sec.~5.2.2) in Ravdess. 
The model also learns associations between the audio and their acoustic properties, which is observed in audio retrieval experiment (Sec.~5.3). Precision$@K$ improves significantly when the model is trained using the acoustic prompts. 



\begin{figure}
    \centering
    \includegraphics[width=\linewidth,height=0.4\linewidth]{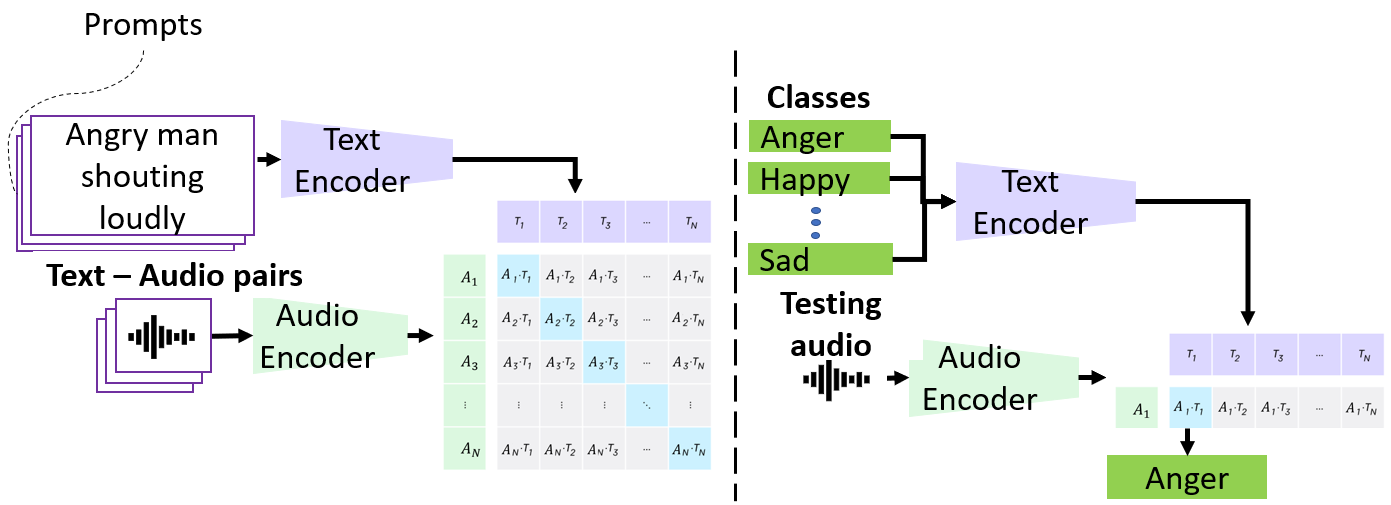}
    \caption{The left part of the image shows the model training. Given a batch of $N$ audio-text pairs, the model trains the audio and text encoders to learn their (dis)similarity using contrastive learning. On the right side is shown an evaluation scenario (leave one out - see Section 5.2.1). Given an audio of unknown emotion, trained audio and text encoders are used to extract representations from the audio and the descriptions. The prediction is made based on the cosine similarity between the two representations.  \vspace{-0.08in}}
    \label{fig:clap_model}
\end{figure}


\vspace{-0.08in}
\section{Contrastive Language-Audio Pretraining}
\vspace{-0.3cm}

Fig.~\ref{fig:clap_model} (left) shows the Contrastive Language-Audio Pretraining (CLAP) model. The audio-text pairs are passed through an audio encoder and a text encoder respectively. Let $f_a(.)$ represent the audio encoder and $f_t(.)$ represent the text encoder. For a batch of N:
\begin{equation}
    \hat{X}_a = f_a(X_a); 
    \hat{X}_t = f_t(X_t)
\end{equation}
where $\hat{X}_a \in \mathbb{R}^{N \times V}$ are the audio representations of dimensionality $V$, and $\hat{X}_t \in \mathbb{R}^{N \times U}$ are the text representations of dimensionality $U$. 

We brought audio and text representations into a joint multimodal space of dimension $d$ by using a projection layer: 
\begin{equation}
    E_a = L_a(X_a);
    E_t = L_t(X_t)
\end{equation}
where $E_a \in \mathbb{R}^{N \times d}$, $E_t \in \mathbb{R}^{N \times d}$, $L_a$ and $L_t$ are the linear projections for audio and text respectively. 

Now that the audio and text embeddings ($E_a$, $E_t$) are comparable, we can measure similarity:
\begin{equation}
    C = \tau*(E_t \cdot E_a^\top)
\end{equation}
where $\tau$ is a temperature parameter to scale the range of logits. The similarity matrix $C \in \mathbb{R}^{N \times N}$ has $N$ correct pairs in the diagonal and $N^2-N$ incorrect pairs in the off-diagonal. The loss can be calculated as: 
\begin{equation}
     \mathcal{L} = 0.5 * (\ell_{text}(C) + \ell_{audio}(C))
\end{equation}
where $\ell_{k} = \frac{1}{N}\sum_{i=0}^{N} \log diag (softmax(C))$ along text and audio axis respectively. We used this symmetric cross-entropy loss ($\mathcal{L}$) over the similarity matrix to jointly train the audio and text encoders along with their linear projections. 


\vspace{-0.1in}
\section{Datasets and CLAP Architecture}
\vspace{-0.3cm}

We use 6 Emotion Datasets (ED) for training and testing, see Table~\ref{tab:emotion_dataset}. The literature using these many datasets is rare. The original CLAP model is trained with audio-text pairs sourced from three audio captioning datasets: ClothoV2~\cite{clotho}, AudioCaps~\cite{audiocaps}, MACS~\cite{macs}, and one sound event dataset: FSD50K~\cite{fsd50k}. Altogether are referred as 4D henceforth. 



\begin{table}
\footnotesize
\caption{Details of the 6 emotion datasets used in this paper.}
    \centering
    \begin{tabular}{lccl}
    \toprule
        Dataset &Files&Class&Emotions\\
    \midrule
        \multirow{2}{0pt}{CMU-MOSEI\cite{zadeh2018multimodal}} & \multirow{2}{0pt}{23K}  & \multirow{2}{0pt}{9} & ang, exc, fear, sad, frus, neu  \\ 
        & & &  surprise, hap, disappoint\\
        \hline
        \multirow{2}{0pt}{IEMOCAP\cite{busso2008iemocap}} & \multirow{2}{0pt}{10K}& \multirow{2}{0pt}{9} & hap, fear, sad,  surprise, exc, \\ 
        & & &  ang, neu, disappoint, frus \\
        \hline
        \multirow{2}{0pt}{MELD\cite{poria2018meld}} & \multirow{2}{0pt}{10K}  & \multirow{2}{0pt}{7} & neu, surprise, fear, sad, \\ 
        & & &  joy, disgust, ang \\
        \hline
        \multirow{2}{0pt}{CREMA-D\cite{cao2014crema}} & \multirow{2}{0pt}{7K}& \multirow{2}{0pt}{6} & ang, disappoint, fear, hap, \\
        & & &  neu, sad \\
        \hline
        \multirow{2}{0pt}{RAVDESS\cite{livingstone2018ryerson}}  & \multirow{2}{0pt}{2.5K}& \multirow{2}{0pt}{8} & neu, calm, hap, sad, \\ 
        & & &  ang, fear, disgust, surprise \\
        \hline
        CMU MOSI\cite{zadeh2018multimodal} & $\qquad$ 2.2K  &   $\mkern7mu$ 3 & neu, positive, negative \\
    \bottomrule
    \end{tabular}
    \label{tab:emotion_dataset}
\end{table}


The architecture is based on the CLAP model in \cite{elizalde2022clap}. We chose this architecture because it yields SoTA performance in learning audio concepts with natural language description. We use log Mel spectrograms from the audios, sampled at 44K Hz, as input to the audio encoder - CNN14~\cite{kong2020panns}, which is pretrained on 2M audio clips from AudioSet. The text encoder is BERT uncased. The audio encodings are of 1024 dimensional
whereas text encodings are 768 dimensional. Both encodings are then projected into a joint multimodal space of dimension 1024. 
Both audio and text encoders are frozen in our experiments, but the projection layers are learnable. 
We use PyTorch to implement the model architecture. The model is trained with 0.0001 learning rate, batch size of 128, for 30 epochs using Adam optimizer. 
\vspace{-0.05in}

\vspace{-0.05in}

\vspace{-0.1in}
\section{Prompt Generation}
\vspace{-0.3cm}

Emotion datasets do not have associated descriptions for each audio. Therefore, we devise a scalable and automatic prompting method that is based on the acoustic properties of the speech audios. There are numerous acoustic properties that describe emotions, as discussed in Section 1. Hence, we hypothesize that including that information in the prompts would benefit emotion recognition. We calculate the pitch and intensity using Librosa \cite{mcfee2015librosa} and we calculate speech rate and articulation rate using Praat \cite{praat}. We construct the prompts in the manner described below: 


\vspace{-0.15 in}
\subsection{Class label (Prompt)}
\vspace{-0.1in}
The simplest description for each audio can be the class label, i.e. `anger'. Thus, we use this as the baseline prompt to compare against the proposed prompts. 

\vspace{-0.15 in}
\subsection{Pitch Prompt}
\vspace{-0.1in}
Pitch is known to be affected by emotion, lower pitch is related with negative emotions like fear and high pitch is related with positive emotions like happiness or surprise \cite{scherer1972acoustic}. Since pitch is naturally sex specific, we bin the average pitch into both based on sex and without sex. Including sex information, we bin the average pitch into four classes, low-male pitch ($<132.5$ Hz), high-male pitch ($>132.5$ Hz,  $<180$ Hz), low-female pitch ($>180$ Hz, $<210$ Hz) and high-female pitch ($>210$ Hz). For the sex agnostic case, we bin into two categories based on cutoff of $170$ Hz. The cutoffs are obtained from the average numbers for vocal pitch reported in literature. The prompt is set as `{bin-class} {emotion-class}', an example of which is `low pitch anger' (without sex information) or `low male pitch anger' (otherwise). \\
\vspace{-0.35in}
\subsection{Intensity Prompt}
\vspace{-0.05in}
Intensity is known to be affected by emotion, low intensity is linked with negative emotion like sadness or melancholy and high intensity is linked with joy or excitement \cite{scherer1972acoustic}. We bin the average intensity over the audio clip in two bins, low and high intensity at $60$ dB.
The same rule as pitch prompt is followed to form the intensity prompt, an example of which is `high intensity anger'.\\
\vspace{-0.35in}
\subsection{Speech-rate Prompt}
\vspace{-0.05in}
It has been observed that faster spoken speech is linked with highly potent emotions such as anger and happiness whilst slower speech is linked with sadness, disgust and boredom \cite{frick1985communicating}. Speech rate is calculated by extracting the number of syllables spoken divided by the total duration of the audio clip. We use $3.12$ syllables/sec as the cutoff to bin the speech rate into two bins, low and high speech rate. An example of speech-rate prompt is `high speech rate anger'.\\
\vspace{-0.35in}
\subsection{Articulation-rate Prompt}
\vspace{-0.05in}
Similarly to speech rate, fast articulation rate is linked with emotions of interest, fear or happiness; whereas slow articulation rate is indicative of sadness and disgust \cite{frick1985communicating}. Articulation rate is calculated as total number of syllables divided by the total phonation time. We bin the audio into low and high articulation rate at the cutoff of $4$ syllables/sec.
An example of articulation-rate prompt is `high articulation rate anger'. \\ 
Even though speech and articulation rate are similar concepts, speech rate captures speaker specific information in the form of number of pauses and hesitation whereas articulation rate would ignore such information.
\vspace{-0.2in}
\subsection{Prompt Augmentation}
\vspace{-0.05in}
To combine all 5 prompts, we pair an audio clip independently with each acoustic prompt. Thus, one audio clip will result in 5 pairs used for training our model.  


\vspace{-0.15in}

\section{Experiments and Results}

\subsection{Prompt Analysis}
\vspace{-0.05in}
To evaluate which of the proposed acoustic prompts are better and to access if any of them are better than the class label, we apply the trained model on emotion classification. 
The model is trained 6 different times, where each time the description associated with emotion audios are varied. Among the 6, 1 is using the class label alone, 4 are using the acoustic prompts as described in Section 4, and 1 is using the prompt augmentation - which combines all the acoustic prompts.

\begin{figure}[tbh!]
  \centering
  \vspace{-0.3cm}
  \includegraphics[width=0.45\linewidth,height=0.25\linewidth]{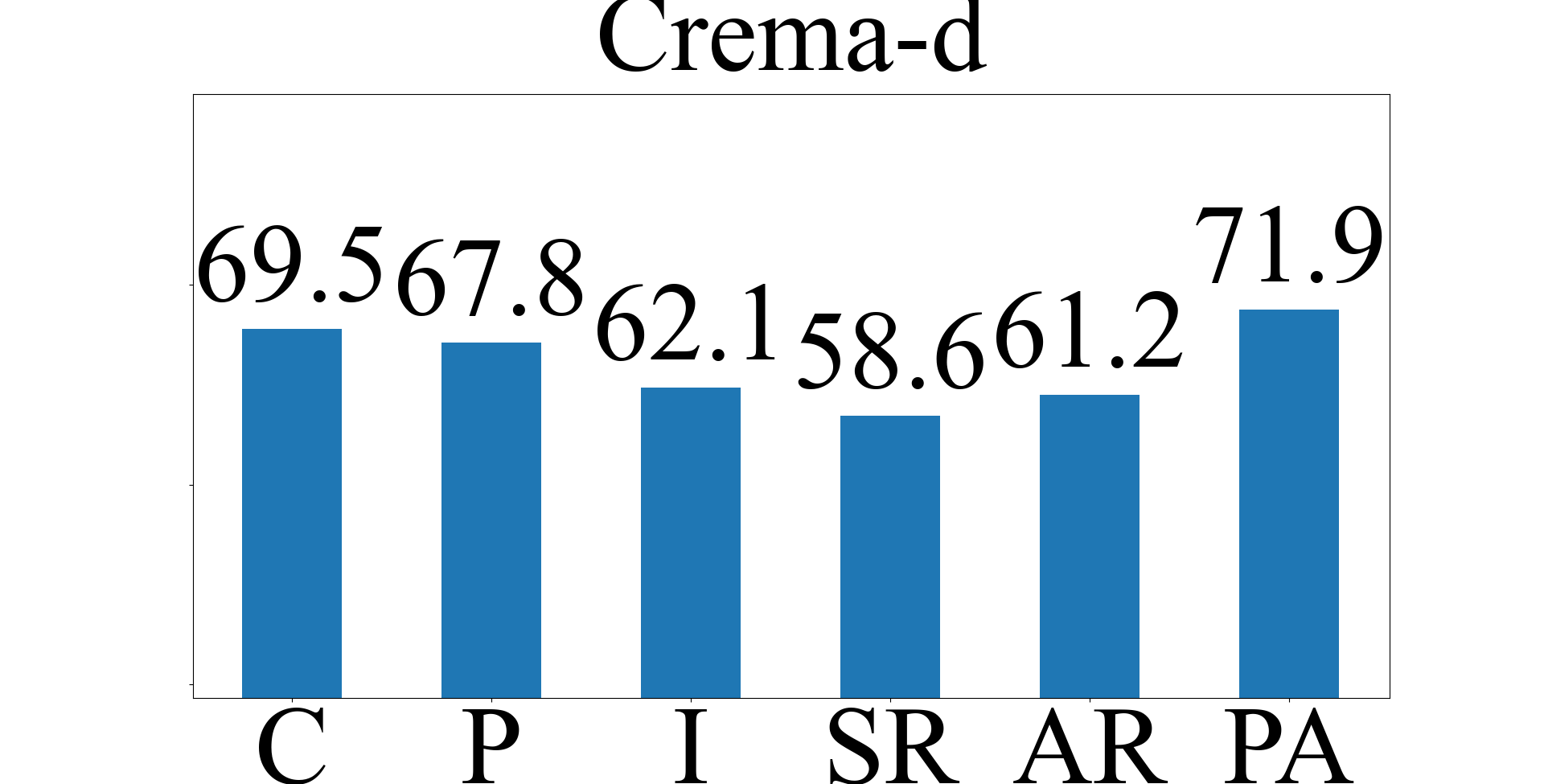}
  \includegraphics[width=0.45\linewidth,height=0.25\linewidth]{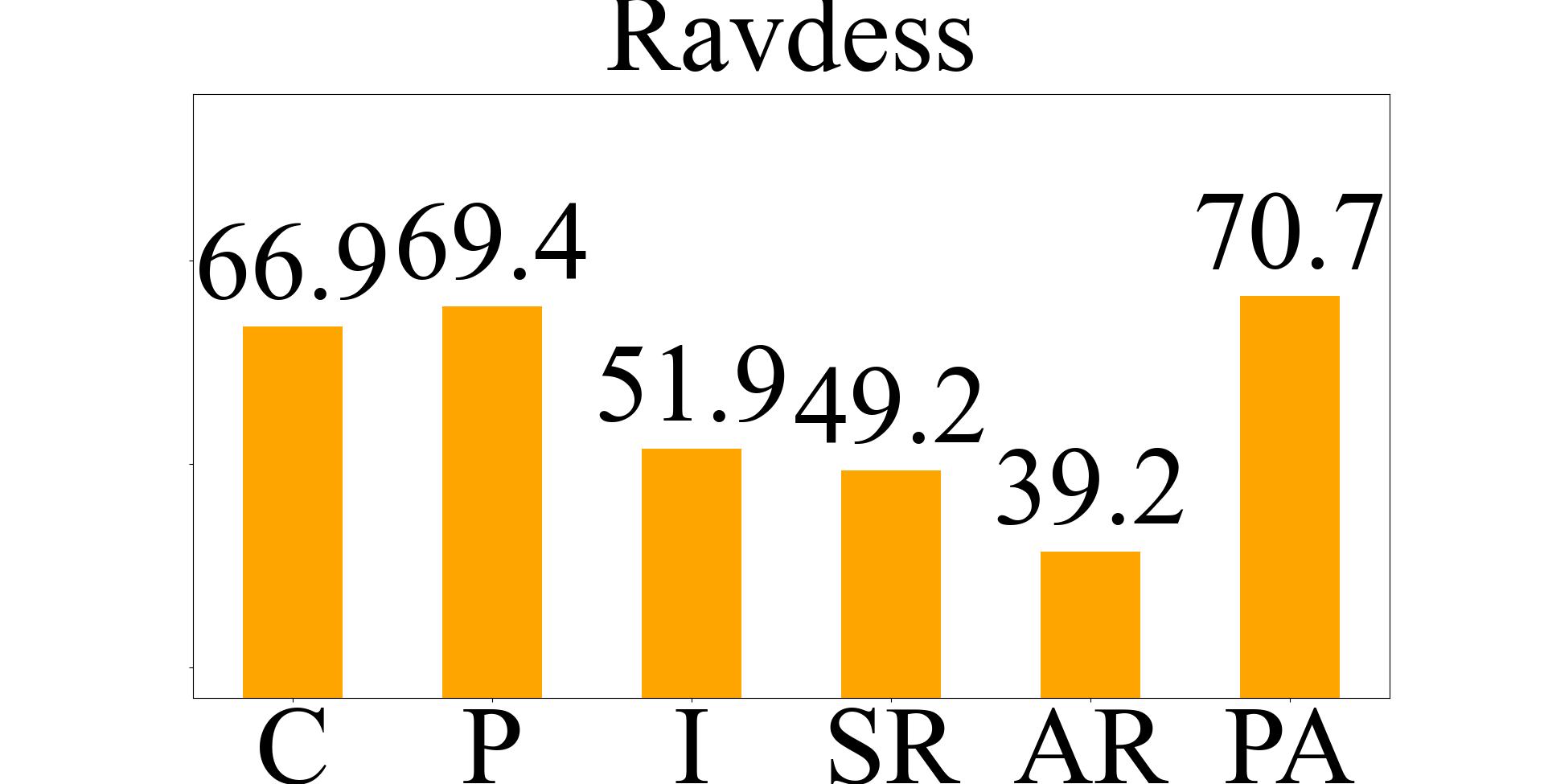}
  \caption{Accuracy achieved using different prompts on Crema-d (left) and Ravdess (right). C=Class label, P=Pitch prompt, I=Intensity prompt, SR=Speech-Rate prompt, AR=Articulation-Rate prompt, PA=Prompt Augmentation.}
    \label{fig:prompt_analysis}
          \vspace{-0.1in}
\end{figure}
We train the model on 4 audio captioning datasets and 1 emotion dataset. The left pat of Figure \ref{fig:prompt_analysis} shows the performance achieved when the model is trained on the training set (including 4D and Ravdess) and tested on the testing set of Ravdess. When done similarly for crema-d, the performance achieved is shown on the right side of Figure \ref{fig:prompt_analysis}.
We observe that among the 4 acoustic prompts, pitch prompt gives the best performance. Next best performance is achieved by the intensity prompt. This can be observed in both Ravdess and Crema-d. On Crema-d, articulation rate prompt performs better than speech rate prompt but the reverse is observed in Ravdess.   
Secondly, we observe that overall prompt augmentation is giving the best performance in both datasets. This validates our original hypothesis that using acoustic prompts would help the emotion classification.

\vspace{-0.1cm}
\begin{table*}[tbh!]
\caption{Accuracy \% when the model is trained under different settings and tested on the left out dataset - shown in the  col 2,3. The same model when finetuned on the left out dataset - result shown in col 4,5. 4D refers to the 4 audio captioning datasets used in original CLAP model. ED refers to Emotion Datasets.}
    \centering
    \begin{tabular}{lcc|cc}
         \toprule
         & \multicolumn{2}{c}{Leave one out dataset} & \multicolumn{2}{c}{Finetune}  \\
         \toprule
         Training dataset & Crema-d & Ravdess & Crema-d & Ravdess \\
         \midrule
         Random & 16.67 & 12.50 & 16.67 & 12.50 \\
         4D & 17.85 &	15.99 &	67.29	& 68.50 \\
        5 ED - \textit{class label} & 	\textbf{35.22}&	22.88 & 68.54 & 68.50 \\
        4D + [5 ED - \textit{class label}] &	34.68 &	\textbf{38.46}  &	\textbf{72.86}&	68.69\\
        4D + [5 ED - \textit{prompt augmentation}] &	33.00 &	27.88 & 72.56	& \textbf{72.46}\\
        SoTA & - & - & 74.70 \cite{wu2022ability} & 81.82 \cite{luna2021proposal} \\
        \bottomrule
        \vspace{-1cm}
    \end{tabular}
    \label{tab:ood_supervised_results}
\end{table*}

\vspace{-0.1in}
\subsection{Emotion Classification}
\vspace{-0.1in}
To evaluate how the acoustic prompts would help in emotion classification, we perform the following two experiments. The first is a zero-shot like setup where we leave one dataset out, which is used during the testing stage. The second is a finetuning setup where the model from the first setup is fine tuned on the left out dataset.

\vspace{-0.2in}
\subsubsection{Leave one out}
\vspace{-0.1in}
This setup evaluates how well a model trained on a pre-defined set of classes generalizes to a new dataset, which might have same or different sets of classes. 
Out of the 6 emotion datasets, we leave one out for testing and train the model on the other 5 emotion datasets. Therefore the training and testing datasets are completely different. In case where Ravdess is the testing dataset, `calm' class is not represented in any of the other training datasets and is a zero shot classification result. In case of Crema-d, all the emotion classes are represented in the training classes from other datasets. 

We perform 4 experiments, shown in columns 2,3 of Table~\ref{tab:ood_supervised_results}. 
There are two main takeaways from this experiment. Firstly adding Emotion datasets in the training stage is helping the performance on the left out emotion dataset. This can be observed in both Ravdess and Crema-d. For crema-d the performance improves from $17.85\%$ to $35.22\%$ just by changing from 4D to 5ED in the training sets. In ravdess, the performance improves from $15.99\%$ to $22.88\%$. 

Secondly, prompt augmentation's results are similar to the results obtained when trained using only the class label. We believe this is happening because of the distribution shift in the training and testing datasets. This also effects distributions of the acoustics - which directly effects the acoustic prompts in the training and testing datasets. For example, `high intensity anger' might not be occurring in the training datasets, but is present in the testing dataset.
This is harming the benefit learnt from the acoustic prompts to be transferable to a completely new dataset at testing time. 
However we do observe improvement using acoustic prompts when we perform the finetuning experiment in the next section. 
Note that the SoTA performance for this evaluation setup is not found in literature because the general evaluation setup is when the dataset is present in both training and testing sets. 

\vspace{-0.15in}
\subsubsection{Finetune}
\vspace{-0.1in}
In this experiment, we want to analyse how using the acoustic prompts at pre-training stage would benefit the classification. We take the model from the previous stage and perform fine tuning on the dataset that had been left out. 

The results for supervised classification are shown in columns 4,5 of Table~\ref{tab:ood_supervised_results}. 
The main takeaway from this experiment is that with finetuning, prompt augmentation shows improvement over the other models. In Ravdess, we see improvement in performance by absolute $3.77\%$, from $68.69\%$ to $72.46\%$. In Crema-d the performance is about the same as when the prompt augmentation is not used - $72.86\%$ vs $72.56\%$, however not significantly different. 

\vspace{-0.15in}
\subsection{Emotion Audio Retrieval}
\vspace{-0.1in}
With the increasing sizes of the audio databases, to be able to search such databases for specific types of audios is important. Therefore we evaluate our models specifically for the audio retrieval task. This would access whether the trained model learns the associations between the acoustic prompts and the respective acoustic properties. We make queries similarly to the prompts as described in Section 4. 
For a given query, the model outputs top $K$ audios whose audio embeddings have highest cosine similarity to the text embedding of the query. For the top $K$ audios, since we know the true acoustic prompts, we can evaluate how good the model's outputs is w.r.t the query. 

\begin{figure}[h!]
    \centering
    \includegraphics[width=0.9\linewidth,height=0.49\linewidth]{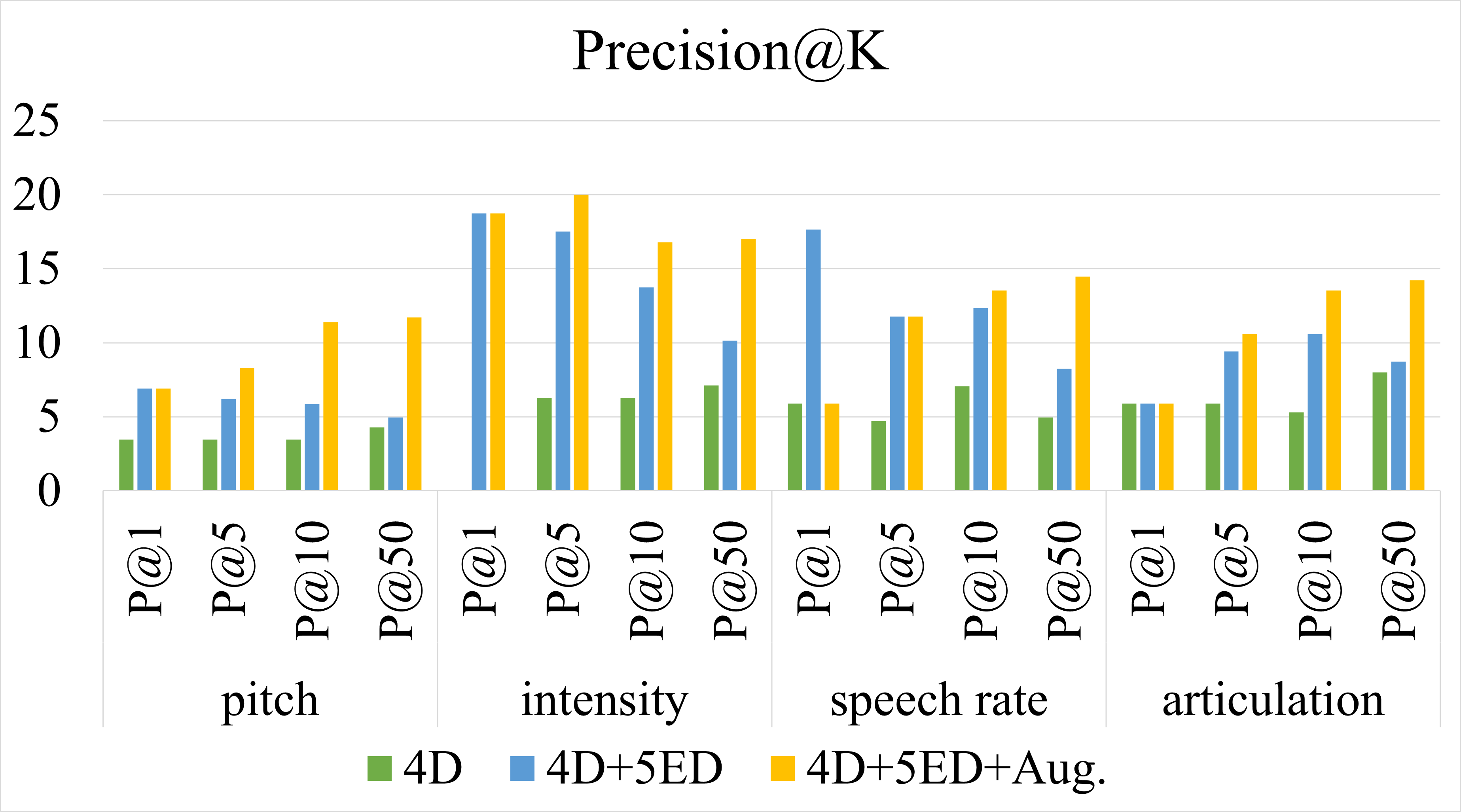}
    \caption{Precision@K achieved when the trained model under different settings is used for the audio retrieval task.}
    \label{fig:retrieval}
\end{figure}

Figure~\ref{fig:retrieval} shows the results of audio retrieval. We calculate precision@$K$ for each acoustic prompt shown on the x-axis. 
From the results we observe that the model trained on all datasets, and using prompt augmentation performs the best in all cases (all types of queries). The takeaway here is that our model is able to retrieve audio significantly better when trained using prompt augmentation. The precision@$K$ numbers are comparable to numbers observed in audio retrieval tasks \cite{kim2019improving}. 
The results are encouraging since this suggests that we can introduce even more elaborate descriptions for each audio and the model will learn associations and be able to retrieve audios with those descriptions. 


\vspace{-0.3in}
\section{Conclusion}
\vspace{-0.3cm}
This work performs emotion recognition using the audios and their natural language descriptions. We use the acoustics of emotions to prompt the audios, in fact there can be more complicated descriptions, invoking the semantics, environment, context among other factors. We envision that as methods of describing emotions become more complicated, our ability to model more nuanced emotions will become better. 
The acoustic properties we extract include pitch, intensity, speech rate and articulation rate from the audios.
We find that among the acoustic prompts, pitch prompt is the best performing. Overall, when we do prompt augmentation it achieves the highest accuracy and improves the performance in Ravdess by $3.8\%$.
We also perform emotion audio retrieval and find that when using model trained on prompt augmentation, we get the significantly better retrieval performance. 
 


\vfill\pagebreak

\bibliographystyle{IEEEbib}
\bibliography{IEEE}

\end{document}